\documentclass[prd,twocolumn,superscriptaddress,nofootinbib,preprintnumbers]{revtex4}
\usepackage[latin1]{inputenc}
\usepackage{graphicx}
\usepackage{amssymb}

\newcommand{\bea}{\begin{eqnarray}}
\newcommand{\ena}{\end{eqnarray}}

\newcommand{\rd}{{\rm d}}

\newcommand{\pa}{\partial}

\def\Vec#1{\mbox{\boldmath $#1$}}

\def\Lap{{\mathop{\Delta}\limits^{(3)}}}

\begin{document}

\title{Generic estimates for magnetic fields generated during inflation \\
including Dirac-Born-Infeld theories}

\author{Kazuharu Bamba}
\affiliation{
Department of Physics, National Tsing Hua University, Hsinchu, Taiwan 300}

\author{Nobuyoshi Ohta}
\affiliation{
Department of Physics, Kinki University, Higashi-Osaka, Osaka 577-8502, Japan}

\author{Shinji Tsujikawa}
\affiliation{
Department of Physics, Faculty of Science, Tokyo University of Science,
1-3, Kagurazaka, Shinjuku-ku, Tokyo 162-8601, Japan}



\begin{abstract}

We estimate the strength of large-scale magnetic fields produced
during inflation in the framework of Dirac-Born-Infeld (DBI) theories.
This analysis is sufficiently general in the sense that
it covers most of conformal symmetry breaking theories in which
the electromagnetic field is coupled to a scalar field.
In DBI theories there is an additional factor associated with the speed
of sound, which allows a possibility to lead to an extra amplification
of the magnetic field in a ultra-relativistic region.
We clarify the conditions under which
seed magnetic fields to feed the galactic dynamo mechanism
at a decoupling epoch as well as present magnetic fields
on galactic scales are sufficiently generated
to satisfy observational bounds.

\end{abstract}


\preprint{KU-TP 022}

\maketitle


\section{Introduction}

It is observationally known that there exist magnetic fields
in clusters of galaxies with the field strength
$10^{-7}-10^{-6}$\,G on 10\,kpc$-$1\,Mpc scales~\cite{Kim1}
as well as those with the field strength $\sim 10^{-6}$\,G
on $1-10$\,kpc scales in galaxies of all types \cite{review}
and in galaxies at cosmological distances~\cite{Kronberg}.
In particular, it is very mysterious that magnetic fields in clusters of
galaxies are as strong as galactic ones and that the coherence
scale may be as large as $\sim$Mpc.
Although galactic dynamo mechanisms~\cite{EParker} have been proposed
to amplify very weak seed magnetic fields up to $\sim 10^{-6}$G,
seed magnetic fields to feed on is necessary at initial stage,
and the effectiveness of the dynamo amplification mechanism
in galaxies at high redshifts and clusters of galaxies is not well
established yet.

Proposed scenarios for the origin of cosmic magnetic fields fall
into two broad categories.
One is astrophysical processes~\cite{Biermann},
and the other is cosmological processes,
e.g., cosmological phase transition~\cite{Baym},
and primordial density perturbations before
the epoch of recombination~\cite{Mata}.
It is difficult, however, that these processes generate
magnetic fields on megaparsec scales with sufficient strength
consistent with observations of galaxies and clusters of galaxies
without dynamo amplification mechanism.

The most natural origin of such a large-scale magnetic field is
electromagnetic quantum fluctuations generated at the inflationary
stage~\cite{Turner},
because inflation has a causal mechanism to generate super-Hubble
gauge fields from microphysical processes.
When we assume the Friedmann-Robertson-Walker (FRW) spacetime
usually considered, its metric is conformally flat.
Moreover, the classical electrodynamics is conformally invariant.
Hence, the conformal invariance of the Maxwell theory must have been
broken at the inflationary stage in order that electromagnetic quantum
fluctuations can be generated at that time~\cite{Parker}.
We note that this does not apply when the FRW background has nonzero spatial
curvature~\cite{NZC}.
(In Refs.~\cite{Maroto}, the breaking of conformal flatness of the
FRW metric induced by the evolution of scalar metric perturbations at
the end of inflation was discussed.
Moreover, the generation of magnetic fields from grand unified theories
(GUT) was studied in Ref.~\cite{Berera:1998hv}.)

So far various conformal symmetry breaking mechanisms have been proposed.
An incomplete list includes: non-minimal gravitational
coupling~\cite{Turner,nonminimal},
dilaton electromagnetism~\cite{dilaton}, coupling to a scalar field \cite{scoupling},
that to a pseudoscalar field \cite{pseudoscalar},
that to a charged scalar field~\cite{charged}, scalar electrodynamics~\cite{scalarele},
general coupling to a time-dependent background field \cite{BS,BS2},
the photon-graviphoton mixing~\cite{Gasperini00},
conformal anomaly induced by quantum effects~\cite{Dolgov},
spontaneous breaking of the Lorentz invariance~\cite{Ber98}
(see also~\cite{LV2}),
the generation of the mass of the gauge field due to a
minimally supersymmetric standard model flat direction
condensate~\cite{Enq},
the photon mass generation due to the existence of the minimal
fundamental scale \cite{cutoff}, nonlinear electrodynamics \cite{NE},
and cosmic defects~\cite{Hollenstein:2007kg}.

In addition, as a breaking scenario based on the fundamental theory of
particle physics, there exists a scenario in the framework of
the Dirac-Born-Infeld (DBI) theory, which is a four
dimensional low-energy effective theory of
string theories~\cite{Aha,Sen,Silverstein}.
In this paper we shall derive the equation of electromagnetic fields
for such theory and estimate the strength of magnetic fields
generated during inflation.
As we will see later, this analysis also covers theories
that possess electromagnetic couplings of the form
$I(\phi, R)F_{\mu \nu}F^{\mu \nu}$~\cite{BS},
where $I$ is an arbitrary function of a scalar field $\phi$ or a Ricci scalar $R$.
Thus the strength of magnetic fields we will derive in this paper is
applicable to many conformal symmetry violating models.
In fact we shall apply our formula to several concrete models of inflation.

This paper is organized as follows.
In Sec.~II we consider the evolution of the $U(1)$ gauge field and derive
the general formula for the field strength of the large-scale magnetic fields.
We apply the derived formula to several inflation models
in Sec.~III.
Finally, Sec.~IV is devoted to conclusions.

We use units in which $k_\mathrm{B} = c = \hbar = 1$,
and adopt Heaviside-Lorentz units in terms of electromagnetism.

\section{Generation of magnetic fields}

Let us start with the following 4-dimensional action
\begin{eqnarray}
\label{DBI}
\hspace*{-0.5em}S &=& -\int {\rm d}^{4}x
 f_1(\phi) \sqrt{-{\rm det}
\left(g_{\mu \nu}+f_2(\phi)
\partial_{\mu} \phi \partial_{\nu} \phi
+f_3(\phi)F_{\mu \nu}\right)} \nonumber\\
& &+\tilde{S} (\phi, R, g_{\mu \nu})\,,
\end{eqnarray}
where $f_1(\phi)$, $f_2(\phi)$, $f_3(\phi)$
are the functions of $\phi$, $g_{\mu \nu}$ is
the metric tensor, and
$F_{\mu\nu} = {\partial}_{\mu}A_{\nu} - {\partial}_{\nu}A_{\mu}$
is the electromagnetic field-strength tensor.
The action $\tilde{S}$ depends on $\phi$, $R$
and $g_{\mu \nu}$ but not on $F_{\mu \nu}$.
The DBI scenario proposed in Ref.~\cite{Silverstein}
corresponds to $f_1(\phi)=1/f(\phi)=\phi^4/\lambda$,
$f_2(\phi)=f(\phi)$ and $f_3(\phi)=\sqrt{f(\phi)}$
for the Anti de-Sitter (AdS) throat.
The rolling tachyon scenario \cite{Sen} corresponds to
$f_1(\phi)=V(\phi)$, $f_2(\phi)=1$ and
$f_3(\phi)=2\pi/M_s^2$, where $M_s$ is the
string mass scale.

When the action (\ref{DBI}) is varied with
respect to the $U(1)$ gauge field $A_{\mu}$,
we neglect those terms whose orders are higher than
$F_{\mu \nu}F^{\mu \nu}$. We then obtain
\begin{eqnarray}
\label{Fmunueq}
\partial_{\mu} \left( \frac{f_1(\phi)f_3^2(\phi)}
{\sqrt{-G}} G F^{\mu \nu} \right)=0\,,
\end{eqnarray}
where $G = {\rm det}\,(G_{\mu \nu})$,
$G_{\mu\nu}=g_{\mu\nu}+f_2(\phi)\pa_\mu \phi \pa_\nu \phi$,
and $F^{\mu \nu}=G^{\mu \alpha}G^{\nu \beta}F_{\alpha \beta}$.
Let us consider the flat FRW spacetime with scale factor $a(t)$,
where $t$ is a cosmic time.
For the Coulomb gauge, ${\partial}^jA_j(t,\Vec{x}) =0$
and $A_0(t,\Vec{x}) = 0$, the equation of motion
for $A_i$ is given by
\bea
\label{Aieq}
\ddot{A_i}(t,\Vec{x})
+\frac{\dot{\mathcal{F}}}{\mathcal{F}}\dot{A_i}(t,\Vec{x})
- \frac{1}{\gamma^2} \frac{1}{a^2}\Lap\, A_i(t,\Vec{x}) = 0\,,
\ena
where a dot represents a derivative with respect to $t$ and
\bea
\mathcal{F} \equiv f_1f_3^2a \gamma\,,\quad
\gamma \equiv \left[1-f_2(\phi)\dot{\phi}^2 \right]^{-1/2}\,.
\ena

One can expand the gauge field $A_i$ by using annihilation and creation operators
together with two orthonormal transverse polarization vectors \cite{BS}.
Then the Fourier mode $A(\eta, k)$, with a conformal time $\eta=\int a^{-1} {\rm d}t$
and a comoving wavenumber $k$, satisfies the following equation of motion:
\bea
\label{Aieq2}
\frac{\rd^2}{\rd \eta^2}A(\eta,k)
+\frac{1}{J}
\frac{\rd J}{\rd \eta}
 \frac{\rd}{\rd \eta} A(\eta,k)
+\frac{k^2}{\gamma^2}A(\eta,k)=0\,,
\ena
where $J=f_1f_3^2\gamma$.
Introducing another time $\tau=\int \gamma^{-1}\,{\rm d}\eta$,
Eq.~(\ref{Aieq2}) reduces to
\bea
\label{A}
A''(\tau,k)+\frac{I'}{I}A'(\tau,k)+k^2 A (\tau,k)=0\,,
\ena
where a prime represents a derivative with respect to
$\tau$ and
\bea
I=f_1f_3^2\,.
\ena
If we consider conformal symmetry violating Maxwell theories with
the action
\bea
\label{Sac}
S=- \int {\rm d}^4 x \sqrt{-g}\,\left[ \frac14 I(\phi,R)
F_{\mu \nu}F^{\mu \nu}+{\cal L} (\phi, R, g_{\mu \nu})
\right],
\ena
we get the same form of equation as (\ref{A})
apart from the fact that $\tau$ is replaced by
the conformal time $\eta$ \cite{BS}.

The Hubble parameter, $H=\dot{a}/a$, needs to satisfy
the condition $|\dot{H}/H^2| \ll 1$ during inflation.
Then we have $\tau \simeq -(\gamma aH)^{-1}$
under the condition $|\dot{\gamma}/H\gamma| \ll 1$.
The modes starting from the ``sub-Hubble'' regime ($k \gg \gamma aH$)
enter the ``super-Hubble'' regime ($k \ll \gamma aH$)
at a time $\tau_k$ characterized by the condition
$\tau_k \simeq -1/k$.

The WKB sub-Hubble solution to Eq.~(\ref{A})
is $A_{\rm in}=e^{-ik \tau}/\sqrt{2kI}$,
which approaches the Minkowski vacuum state
in the limit $\tau \to -\infty$.
Meanwhile the super-Hubble solution neglecting
correction terms of the order $k^2$ is given by
$A_{\rm out}=C(k)+D(k) \int_{\tau}^{\tau_R}
{\rm d}\tilde{\tau}/I(\tilde{\tau})$,
where $C(k)$ and $D(k)$ are constants and
$\tau_R$ corresponds to the time at reheating.
Matching these two solutions at time $\tau=\tau_k$
using the junction conditions $A_{\rm out}(\tau_k)
=A_{\rm in}(\tau_k)$ and $A'_{\rm out}(\tau_k)
=A'_{\rm in}(\tau_k)$, the coefficients $C(k)$ and
$D(k)$ are determined accordingly.
Neglecting the decaying mode for $A_{\rm out}$,
we get the late-time solution $|A(\tau,k)|^2= |C(k)|^2$
at the end of inflation:
\begin{equation}
\label{Atau}
|A(\tau,k)|^2=\frac{1}{2kI(\tau_k)}
\left| 1- \left( \frac{I'(\tau_k)}{2kI(\tau_k)}+i
\right)\,k \int_{\tau_k}^{\tau_R}
\frac{I(\tau_k)}{I(\tilde{\tau})} \rd \tilde{\tau}
\right|^2.
\end{equation}

In the following we assume that the energy density
of the field $\phi$ is converted to radiation almost
instantly right after the end of inflation and that
the conductivity $\sigma_c$ of the Universe jumps
to a value much larger than the Hubble rate at reheating.
Then the proper magnetic field,
$B_i^{\rm proper}(t,\Vec{x})=a^{-2}\epsilon_{ij \ell}
\partial_j A_{\ell} (t, \Vec{x})$, evolves as
$B_i^{\rm proper}(t,\Vec{x}) \propto a^{-2}$
in the reheating and subsequent
radiation/matter/dark energy dominated stages.
Taking into account two polarization degrees of freedom,
the spectrum of the magnetic field is given by
\bea
\label{Bpro}
|{B}^{\mathrm{proper}}(\tau,k)|^2
=2\frac{k^2}{a^4}|A(\tau,k)|^2\,.
\ena
The energy density of the magnetic field per unit
logarithmic interval of $k$ is defined by
\begin{eqnarray}
\label{rhoB}
\rho_B(\tau,k) &\equiv&
\frac{1}{2}
\frac{4\pi k^3}{(2\pi)^3}|{B}^{\mathrm{proper}}(\tau,k)|^2 I(\tau)\,.
\end{eqnarray}
Since the radiation density evolves as $\rho_{\gamma}(\tau)=
\rho_{\gamma}(\tau_R) (a_R/a)^4$, it is convenient to introduce
the density parameter
$\Omega_B(\tau,k)=\rho_B(\tau,k)/\rho_{\gamma}(\tau)$.
{}From Eqs.~(\ref{Atau}), (\ref{Bpro}) and (\ref{rhoB}) we obtain
\bea
\hspace*{-1.5em}\Omega_B(\tau,k)
&=& \frac{15}{2\pi^4N_{\rm eff}}
\left( \frac{k}{a_R T_R} \right)^4 \frac{I(\tau)}{I(\tau_k)} \nonumber \\
\hspace*{-1.5em}&\times& \left| 1- \left( \frac{I'(\tau_k)}{2kI(\tau_k)}+i
\right)\,k \int_{\tau_k}^{\tau_R}
\frac{I(\tau_k)}{I(\tilde{\tau})} \rd \tilde{\tau}
\right|^2.
\ena
Here we used $\rho_{\gamma}(\tau_R)=\pi^2 N_{\rm eff}T_R^4/30$,
where $N_{\rm eff}$ is the effective massless degree of freedom
and $T_R$ is the reheating temperature.

In order to estimate the strength of magnetic fields,
let us consider the case in which the evolution of
the quantity $I$ during inflation is given by
\bea
\label{I}
I=I_* (\tau/\tau_*)^{-\alpha}\,,
\ena
where $I_*$, $\tau_*$ and $\alpha$ are constants.
This choice is made to get quantitative estimate of the generated magnetic field,
and is general enough to cover many models including those discussed
in the following section.
On using the relations $\tau_R \simeq  -(\gamma_R a_RH_R)^{-1}$
and $3H_R^2 \simeq \rho_{\gamma} (\tau_R)/M_{\rm pl}^2$
(where $M_{\rm pl}$ is a reduced Planck mass), we get
\begin{equation}
\hspace*{-0.6em}\Omega_B (\tau,k)=
{\cal C} \frac{N_{\rm eff}}{1080}
\left( \frac{T_R}{M_{\rm pl}} \right)^4
\left( \frac{k}{a_RH_R} \right)^{4-\alpha}
\frac{I(\tau)}{I(\tau_R)} \gamma_R^{\alpha}\,,
\end{equation}
where ${\cal C}=|1-\frac{\alpha+2i}{2(\alpha+1)}|^2$.
Hence the spectral index of the magnetic field
is given by
\bea
\label{nB}
n_B=4-\alpha\,.
\ena
For larger positive $\alpha$ it is possible to generate
large-scale magnetic fields.
Note that the reheating temperature generally has an upper
bound from the Cosmic Microwave Background (CMB)
observations ($T_R \lesssim 10^{15}$\,GeV).
Because of the presence of the $\gamma$ factor there is
an extra amplification of the magnetic field for
$\gamma_R \gg 1$ and $\alpha>0$.

Let us first estimate the quantity $k/a_RH_R$ for the scale
$L=2\pi/k$\,[Mpc].
Using the present value $H_0^{-1}=3.0 \times 10^{3}\,h^{-1}$\,Mpc
and the relation $a_0/a_R=T_R/T_0$
we have $k/a_RH_R \simeq (1.88/h) (10^4\,{\rm Mpc}/L)(T_R/T_0)(H_0/H_R)$.
Since $H_R^2 \simeq \pi^2 N_{\rm eff} T_R^4/90M_{\rm pl}^2$,
$T_0=2.73$\,K and  $H_0=2.47h \times 10^{-29}$\,K, we find
\bea
\frac{k}{a_RH_R}=5.1 \times 10^{-25}
\frac{1}{\sqrt{N_{\rm eff}}}
\frac{M_{\rm pl}}{T_R}
\frac{1}{L/{\rm Mpc}}\,.
\ena
The energy density $\rho_B (\tau_0)$ at the present epoch
is given by $\rho_B (\tau_0)=(1/2)|B (\tau_0)|^2=
\Omega_B (\tau_0,k)\,\rho_{\gamma}(\tau_0)$, where
$B (\tau_0)$ is an observed magnetic field.
Since $\rho_{\gamma}(\tau_0) \simeq 2 \times 10^{-51}$\,GeV$^4$
and $1\,{\rm G}=1.95 \times 10^{-20}$\,GeV$^2$, we obtain
\bea
\label{Btau1}
|B(\tau_0)| &=& 2.7 \times 10^{-56+25\alpha/2} \cdot
\left[ {\cal C} \frac{I(\tau_0)}{I(\tau_R)} \right]^{1/2}
N_{\rm eff}^{\alpha/4-1/2} \nonumber \\
& & \times \left( \frac{1}{5.1} \frac{T_R}{M_{\rm pl}}
\gamma_R \right)^{\alpha/2}
\left( \frac{L}{{\rm Mpc}} \right)^{\alpha/2-2}
~~{\rm G}\,.
\ena
If we take the maximum reheating temperature $T_R \simeq 10^{15}\,{\rm GeV}
=4 \times 10^{-4}M_{\rm pl}$ with $N_{\rm eff}=100$,
one can estimate the order of the present magnetic field to be
\begin{equation}
\label{Btau2}
|B(\tau_0)| \simeq
10^{11\alpha-57}
\left[ {\cal C} \frac{I(\tau_0)}{I(\tau_R)} \right]^{1/2}
\gamma_R^{\alpha/2}
\left( \frac{L}{{\rm Mpc}} \right)^{\alpha/2-2}
\,{\rm G}\,.
\end{equation}
We must have $|B(\tau_0)| \gtrsim 10^{-9}$\,GeV to explain
observed magnetic fields on the scales 1\,kpc$-$1\,Mpc
without the mechanism of galactic dynamo.

At the decoupling epoch with $z=1000$,
the radiation energy density is given by
$\rho_{\gamma}(\tau_{\rm dec}) \simeq 10^{12} \rho_{\gamma} (\tau_0)$.
Then the magnetic field strength at this epoch is given by
\bea
\label{Btau3}
|B(\tau_{\rm dec})| &=& 2.7 \times 10^{-50+25\alpha/2} \cdot
\left[ {\cal C} \frac{I(\tau_{\rm dec})}{I(\tau_R)} \right]^{1/2}
N_{\rm eff}^{\alpha/4-1/2} \nonumber \\
& & \times  \left( \frac{1}{5.1} \frac{T_R}{M_{\rm pl}}
\gamma_R \right)^{\alpha/2}
\left( \frac{L}{{\rm Mpc}} \right)^{\alpha/2-2}
~~{\rm G}\,.
\ena
When $T_R \simeq 10^{15}\,{\rm GeV}$ and $N_{\rm eff}=100$,
the order of $|B(\tau_{\rm dec})| $ is
\begin{equation}
\label{Btau4}
|B(\tau_{\rm dec})| \simeq
10^{11\alpha-51}
\left[ {\cal C} \frac{I(\tau_{\rm dec})}{I(\tau_R)} \right]^{1/2}
\gamma_R^{\alpha/2}
\left( \frac{L}{{\rm Mpc}} \right)^{\alpha/2-2}
\,{\rm G}\,.
\end{equation}
The seed field with an amplitude
$|B(\tau_{\rm dec})| \gtrsim 10^{-23}$\,G is
required to explain the present size of the magnetic
field through the galactic dynamo mechanism
for a flat universe without cosmological constant.
However this limit is relaxed up to
$|B(\tau_{\rm dec})| \gtrsim 10^{-30}$\,G
on $\sim$kpc scale in the presence of cosmological
constant at late times \cite{Anne}.

We would like to stress here that the above results are valid
even for the theories with the action (\ref{Sac})
by setting $\gamma=1$.

\section{Application to concrete models}

We shall apply the formula derived
in the previous section to several conformal
symmetry breaking models.
We adopt the reheating temperature $T_R=10^{15}$\,GeV
to estimate the maximum allowed size of
magnetic fields.
Note that the factor ${\cal C}$ in
Eqs.~(\ref{Btau1})-(\ref{Btau4}) is of the order of unity.

\subsection{Power-law inflation with $\gamma=1$}

Let us consider the dilatonic coupling $I(\phi)=e^{\lambda \phi}$
and the Lagrangian ${\cal L}=(1/2)(\nabla \phi)^2+V(\phi)$
in Eq.~(\ref{Sac}). This corresponds to the case $\gamma=1$,
i.e., $\tau=\eta$.
If the potential is given by $V(\phi)=V_0 \exp(-\sqrt{2/p}\,\phi)$,
where $\phi$ is normalized by $M_{\rm pl}$,
power-law inflation with $a \propto t^p$ ($p>1$)
is realized.  Since the field evolves as $\phi=\phi_0+\sqrt{2p}\,\ln\,(t)$,
the coupling $I$ has a time-dependence
$I \propto t^{\lambda \sqrt{2p}} \propto (-\eta)^{-\alpha}$, where
\bea
\alpha=\lambda \frac{\sqrt{2p}}{p-1}\,.
\ena

We shall study the case in which the field $\phi$ is frozen right after the end of
inflation due to the appearance of a potential minimum.
We then have $I(\tau_R)=I(\tau_0)=I(\tau_{\rm dec})$ in
Eqs.~(\ref{Btau2}) and (\ref{Btau4}).
In order to get the present size of magnetic fields
($|B(\tau_0)| \gtrsim 10^{-9}$\, G) on the scale $L=1$\,Mpc
without the mechanism of galactic dynamo, we must have $\alpha>4.4$.
To explain the origin of seed magnetic fields $|B(\tau_{\rm dec})|>10^{-30}$\,G
on the scale $L=1$\,Mpc at the decoupling epoch,
we need $\alpha>1.9$.
This condition is relaxed to $\alpha>1.6$ for
the magnetic fields on the scale $L=1$\,kpc.

The recent Wilkinson Microwave Anisotropy Probe (WMAP) data
of density perturbations
constrains the power $p$ to be $p>70$ at the 95\%
confidence level \cite{Komatsu08}.
We then find that the parameter $\lambda$ must
satisfy at least the relation,
$\lambda>9.4$, from the condition $\alpha>1.6$.

\subsection{Tachyon inflation}

The rolling tachyon scenario \cite{Sen,Mal,Fein,Pad} corresponds
to the choice $f_1(\phi)=V(\phi)$, $f_2(\phi)=1$
and $f_3(\phi)=2\pi/M_s^2$, where $M_s$ is the string mass scale.
We then have $I(\phi)=4\pi^2 V(\phi)/M_s^4$,
which decreases during inflation.

Consider the inverse power-law potential $V(\phi)=V_0 \phi^{-2}$
with $V_0=4p(1-2/3p)^{1/2}M_{\rm pl}^2$.
This leads to the power-law expansion $a \propto t^p$ ($p \gg 1$)
with $\phi=\sqrt{2/3p}\,t$ \cite{Fein,Pad,Copeland},
in which case $\gamma$ is a constant [$\gamma=(1-2/3p)^{-1/2} \simeq 1$].
Hence one has $I \propto t^{-2} \propto (-\tau)^{2/(p-1)}$,
thereby giving
\bea
\alpha=\frac{2}{1-p}<0\,.
\ena
This shows that the spectral index $n_B$ given in
Eq.~(\ref{nB}) is highly blue-tilted.
Hence it is difficult to generate sufficient amounts of
large-scale magnetic fields.
Moreover the quantity $I(\phi)$ ($\propto V(\phi)$) decreases toward zero after inflation
for the standard tachyon models in which the field rolls down
toward infinity. In tachyon inflation the magnetic field
at the present epoch is vanishingly small.

\subsection{DBI inflation}

The DBI inflation for the AdS throat corresponds to the choice
$f_1(\phi)=1/f(\phi)=\phi^4/\lambda$, $f_2(\phi)=f(\phi)$
and $f_3(\phi)=\sqrt{f(\phi)}$ \cite{Silverstein}.
In this case one has $I(\phi)=1$, which means that the generation
of the magnetic field does not occur unlike the results found in Ref.~\cite{Ganjali}.
Since the coupling $f_3(\phi)$ given above is chosen to reproduce
the standard Maxwell Lagrangian in the low-energy regime ($f\dot{\phi}^2 \ll 1$),
the field $\phi$ is not directly coupled to the electromagnetic field.

One may consider a scenario in which the coupling $f_3(\phi)$ takes
a different form in the ultra-relativistic regime
$(\gamma=[1-f\dot{\phi}^2]^{-1/2} \gg 1)$.
For example, let us study the case
\bea
f_3(\phi) \propto \phi^{-n}\,,~~~~
{{\rm i.e.,}}~~~~I \propto \phi^{4-2n}\,.
\ena
For the potential $V(\phi)=(1/2)m^2\phi^2$, inflationary
solutions in the regime $\gamma \gg 1$ are given by
$\phi=\sqrt{\lambda}/t$, $\gamma \simeq
mM_{\rm pl}\sqrt{2\lambda/3}/\phi^2 \propto t^2$ and
$a \propto t^p$, where
$p=\sqrt{\lambda/6}\,(m/M_{\rm pl})$ \cite{Silverstein,Shan}.
Since $t$ has a dependence $t \propto (-\tau)^{-1/(p+1)}$
in this case we get $I \propto (-\tau)^{-\alpha}$ with
\bea
\alpha=\frac{2n-4}{p+1}\,.
\ena

For $\gamma \gg 1$ and $\alpha>0$,
the magnetic field can be more significantly
amplified relative to the case $\gamma=1$ because a mode with
the wavenumber $k$ crosses the point $k=\gamma aH$
earlier for larger $\gamma$.
In the ultra-relativistic regime of the DBI inflation the non-Gaussian
parameter $f_{\rm nl}$ in CMB observations is given by
$f_{\rm nl}=\frac{35}{108}(\gamma^2-1)$ \cite{Chen}.
Using the latest WMAP bound $|f_{\rm nl}| < 253$
based on the equilateral models \cite{Komatsu08},
we obtain the constraint $\gamma_{\rm CMB}<28$
on the scales relevant CMB anisotropies.
Since $\gamma$ grows as $\gamma \propto a^{2/p}$
during inflation, one can estimate the value $\gamma_R$
to be $\gamma_R=\gamma_{\rm CMB}\,e^{2N/p}$,
where $N$ is the number of e-foldings from the epoch at which
CMB fluctuations are generated
to the end of inflation ($N=50\sim 60$).
In the following we adopt the value $N=55$ for concreteness.

Let us consider the case in which the field $\phi$ is frozen
right after the end of inflation so that $I(\tau_R)$
is the same order as $I(\tau_{\rm dec})$ and $I(\tau_0)$.
On using Eq.~(\ref{Btau2}), we find that the present magnetic field
greater than the order of $10^{-9}$\,G can be obtained for
\bea
\label{con1}
n>2+\frac{2p(p+1)\left[24+\log_{10}\,(L/{\rm Mpc})\right]}
{48+p\left[22+\log_{10}\,(\gamma_{\rm CMB}\cdot
L/{\rm Mpc})\right]}\,.
\ena
{}From Eq.~(\ref{Btau4}) the condition to get the seed
magnetic field larger than the order of $10^{-30}$\,G
is given by
\bea
\label{con2}
n>2+\frac{p(p+1)\left[21+2\log_{10}\,(L/{\rm Mpc})\right]}
{48+p\left[22+\log_{10}\,(\gamma_{\rm CMB}\cdot
L/{\rm Mpc})\right]}\,.
\ena

In the relativistic regime of DBI inflation the tensor-to-scalar ratio
in CMB anisotropies is given by
$r \simeq 16\epsilon/\gamma=(48/\lambda)(M_{\rm pl}/m)^2
(\phi/M_{\rm pl})^2$ (where $\epsilon=-\dot{H}/H^2$ is the
slow-roll parameter). Using the latest WMAP bound
$r<0.2$ \cite{Komatsu08} together with the non-gaussianity bound
$\gamma=mM_{\rm pl}\sqrt{2\lambda/3}/\phi^2<28$, we find
that $\phi_{\rm CMB}$ is bounded from both above and below.
For the consistency of this inequality, we must require that
$\lambda (m/M_{\rm pl})^2>49$, i.e., $p>2.9$.

If we adopt the values $L=1$\,Mpc, $\gamma_{\rm CMB}=28$ and $p=3$
in Eqs.~(\ref{con1}) and (\ref{con2}), then we get the bounds
$n>6.9$ and $n>4.1$, respectively.
The constraint on $n$ is weakened for smaller scales.
For example, when $L=1$\,kpc, $\gamma_{\rm CMB}=28$ and $p=3$,
Eq.~(\ref{con2}) gives the bound $n>3.6$.
Meanwhile the constraint on $n$ tends to be tighter for larger $p$.
Since $\gamma_{\rm CMB}$ is bounded from above
($\gamma_{\rm CMB}<28$), one can not choose arbitrary
large values of $\gamma_{\rm CMB}$ to make the r.h.s. of
Eqs.~(\ref{con1}) and (\ref{con2}) smaller.

\section{Conclusions}

In the present paper, we have studied the generation of large-scale magnetic
fields due to the breaking of the conformal invariance of the electromagnetic
field through its coupling to a scalar field in the framework of DBI theory.
Introducing a time $\tau=\int \gamma^{-1}\,{\rm d}\eta$, the Fourier
component of the gauge field satisfies the equation of motion
(\ref{A}). This is the same form of equation derived for the electromagnetic
coupling given in Eq.~(\ref{Sac}) apart from the difference that
$\tau$ is replaced by conformal time $\eta$ for the action (\ref{Sac}).
Hence our analysis is applicable to many conformal symmetry
breaking Maxwell theories.

By matching two solutions in ``sub-Hubble'' ($k \gg \gamma aH$) and
``super-Hubble'' ($k \ll \gamma aH$) regimes during the inflationary epoch,
the strength of the magnetic field at the end of inflation can be estimated as
Eq.~(\ref{Atau}). Under the assumptions that the energy density of
inflaton is almost instantly converted to radiation after inflation and that
the conductivity during reheating is much higher than the Hubble rate
at that epoch, we derived the size of the magnetic field both
at the present and at the decoupling epoch.
Note that we have not assumed any other mechanisms for the
amplification of the magnetic field.
The results (\ref{Btau1}) and (\ref{Btau3}) are sufficiently general
to cover the theories described by the action (\ref{Sac}).

We applied our formula for three cases: (i) power-law inflation
with $\gamma=1$, (ii) tachyon inflation, and (iii) DBI inflation.
The power $\alpha$ defined in Eq.~(\ref{I}) that
characterizes the evolution of the quantity $I$ during
inflation is important to determine the spectral index
of the magnetic field.
For the theories with $\gamma=1$, it should be generally
required that the spectrum is red-tilted ($\alpha>4$) to realize the
present field strength $|B(\tau_0)|$ larger than $10^{-9}$\,G
on the scales 1\,kpc$-$1\,Mpc.
The constraint on $\alpha$ is not so severe to obtain seed magnetic
fields to feed the galactic dynamo
mechanism ($|B(\tau_{\rm dec})|>10^{-30}\,$G).
In power-law inflation, for example, we found that the constant
$\lambda$ for the electromagnetic
coupling $I(\phi)=e^{\lambda \phi}$ is constrained to be
$\lambda>9.4$ to satisfy the condition required
for the seed field on the scale $L=1$\,kpc ($\alpha>1.6$).

In the theories with $\gamma \neq 1$ there exists an extra
factor $\gamma_R^{\alpha/2}$ that can lead to additional
amplification of the magnetic field.
In tachyon inflation, in addition to the fact that $\gamma_R$ is
very close to 1, the quantity $I(\phi)$ is proportional to
the field potential $V(\phi)$, which decreases during inflation
(i.e., $\alpha<0$). Hence we can not expect the generation of
large-scale magnetic fields consistent with observations.

In DBI inflation, if we wish to reproduce the standard Maxwell
theory in low-energy regimes, we have $f_1(\phi)=1/f(\phi)
=\phi^4/\lambda$ and $f_3(\phi)= \sqrt{f(\phi)}$ in
the action (\ref{DBI}).
This corresponds to the effective coupling
with $I(\phi)=1$, which means that the generation of
magnetic fields can not be expected.
This situation changes if we allow the possibility that the coupling $f_3(\phi)$
takes a different form in the ultra-relativistic regime ($\gamma \gg 1$).
We adopted the coupling of the form $f_3(\phi) \propto \phi^{-n}$
and derived the bounds (\ref{con1}) and (\ref{con2}) to get
observationally required magnetic fields at the present and at the
decoupling epoch. It is worth mentioning that the presence of the
$\gamma_R^{\alpha/2}$ factor leads to the
larger magnetic field relative to the theories with $\gamma=1$.

It will be certainly of interest to apply our formula to many other
conformal symmetry breaking models.
While we have assumed instant reheating with large conductivity,
the details of the reheating process actually depends upon
models of inflation.
It is generally difficult to construct string/brane inflation
models with successful reheating, so we need to wait for
the construction of such viable models to carry out
detailed analysis for the dynamics of magnetic fields
in the reheating phase.

Finally, we remark interesting cosmological effects of
large-scale magnetic fields generated during inflation
on the CMB radiation.
In Ref.~\cite{Caprini:2003vc}, the effect of gravity waves induced by a
possible helicity-component of a primordial magnetic field
on CMB temperature anisotropies and polarization has been considered.
According to it, the effect could be sufficiently large to be
observable if the spectrum of the primordial magnetic field is close to
scale-invariant and if its helical component is stronger than
$\sim 10^{-10}$\,G.
In Ref.~\cite{Caprini:2003vc}, only the tensor mode, whose contribution
is significant for low multipoles ($l<100$), has been considered,
while the vector mode has an imprint for higher multipoles
too~\cite{Kahniashvili:2005xe}.
Thus, the tensor mode alone can not significantly 
limit the magnetic field amplitude.
According to Ref.~\cite{Caprini:2003vc}, the amplitude of the
helical magnetic field (and not the helical component) must be larger than
a few $\times 10^{-9}$\,G to be detectable by current CMB measurements.
Similar bounds have been derived in Ref.~\cite{CMB-Limit}.
However, the future missions, for example, PLANCK, will
be able to test the cosmological magnetic field with an amplitude
$10^{-10}$\,G or even lower~\cite{Kristiansen:2008tx}.
The current (best) limit on the amplitude of the magnetic field from the CMB
polarization Faraday rotation effect using WMAP 5 years data is
around 5 $\times 10^{-10}$\,G~\cite{Kahniashvili:2008hx} for
the magnetic field generated from inflation.


\section*{Acknowledgements}
KB was supported in part by the Open Research Center Project
at Kinki University and
National Tsing Hua University under Grant \#: 97N2309F1.
This work was supported in part by the Grant-in-Aid for
Scientific Research Fund of the JSPS for Scientific Research
Nos.~20540283 and 06042 (NO) and 30318802 (ST),
also in part by the Japan-U.K. Research Cooperative Program.


\end{document}